\documentclass[11pt,preprint,preprintnumbers,amsmath,amssymb,nofootinbib,aps]{revtex4}
\usepackage{graphics,color,array,dcolumn}
\usepackage{hyperref}
\usepackage{calc}

\usepackage{amsmath}
\usepackage{amssymb}
\usepackage{xspace}
\newlength{\figurewidth}
\usepackage{graphicx}
\usepackage{amsfonts}

\allowdisplaybreaks[1]
\def\STr{\mathrm{STr}}
\def\calo{{\cal O}}
\def\calr{{\cal R}}

\begin{document}
\setlength{\figurewidth}{\columnwidth}

\title{On the scheme dependence of gravitational beta functions
\footnote{Proceedings of the 49-th Cracow School of Theoretical Physics, "Non-perturbative Gravity and 
Quantum Chromodynamics", May 31-June 10, 2009.  To appear in Acta Physica Polonica.}
}
\author{Gaurav Narain}
\affiliation{SISSA, Via Beirut 4, 34151 Trieste, Italy}
\email{narain@sissa.it}
\author{Roberto Percacci\footnote{\it on leave from SISSA, via Beirut 4, I-34151 Trieste, Italy. 
Supported in part by INFN, Sezione di Trieste, Italy}}
\email{rpercacci@perimeterinstitute.ca}
\affiliation{Perimeter Institute for Theoretical Physics, 31 Caroline St. North, Waterloo, Ontario N2J 2Y5, 
Canada}


\begin{abstract}
We discuss the arbitrariness in the choice of cutoff scheme
in calculations of beta functions.
We define a class of ``pure'' cutoff schemes, in which the cutoff
is completely independent of the parameters that appear in the action.
In a sense they are at the opposite extreme of the ``spectrally adjusted'' 
cutoffs, which depend on all the parameters that appear in the action.
We compare the results for the beta functions of Newton's constant and 
of the cosmological constant obtained with a typical cutoff
and with a pure cutoff, keeping all else fixed.
We find that the dependence of the fixed point on an arbitrary parameter 
in the pure cutoff is rather mild.
We then show in general that if a spectrally adjusted cutoff produces
a fixed point, there is a corresponding pure cutoff that 
will give a fixed point in the same position.
\end{abstract}
\maketitle


\section{Introduction}

In the last ten years much work has been done on the calculation of
the beta functions of gravity, using Wilsonian Renormalization Group methods.
Such calculations have focused first on the Hilbert action,
usually including also a cosmological term \cite{Dou, Reuter, Souma, Lauscher, Litim, Eichhorn},
and have been extended to include terms quadratic in curvature
\cite{Lauscher2, Codello, BMS, Niedermaier:2009zz}
as well as higher polynomials in the Ricci scalar \cite{CPR1, CPR2, MachSau}
or in some approximation even all orders of the derivative expansion \cite{PercacciN}.
Similar studies have also been done in the conformal reduction of gravity,
where only the conformal factor is dynamical \cite{conformal}.
The main motivation for this work has been to try to (dis)prove the asymptotic safety of gravity,
{\it i.e.} the existence of a fixed point with finitely many UV attractive directions
\cite{Weinberg}, also \cite{AS_rev} for reviews.

One slightly bothering aspect of these calculations is that very often one
works with quantities that are scheme-dependent, {\it i.e.} depend on details of how
one chooses to implement a cutoff.
Scheme dependence is nothing new in quantum field theory.
It is usually taken as a sign that the quantity one is calculating is not
directly measurable.
The leading approximation to the beta functions of dimensionless coupling constants,
such as Yang-Mills and Yukawa couplings in four dimensions is scheme-independent.
In the case of gravity the most interesting couplings, like
Newton's constant and the cosmological constant are dimensionful
and their beta functions are scheme dependent even in the lowest approximation.
One hopes that at least the qualitative features of the Renormalization Group (RG) flow
are scheme independent. 
For example, the existence of a Fixed Point (FP) 
as well as the number of attractive/repulsive directions at the FP should be scheme-independent.
More quantitatively, one expects also the critical exponents at the FP to be scheme-independent.
All the calculations performed so far support the existence of a FP with finitely many
attractive directions. For pure gravity the number of attractive directions seems likely to be three
\cite{CPR1, CPR2, MachSau, BMS}, in accordance with the idea of asymptotic safety.
At a quantitative level, when using a truncation, 
both the position of the FP and the critical exponents depend on the
scheme, but there are some quantities, such as the dimensionless combination $\Lambda G$ which are much 
less sensitive to cutoff scheme than others.

Here we add some tassels to this picture. We consider a class of cutoff schemes
that have not received much attention in the literature, that we call ``pure''.
We say that a cutoff is ``pure'' if it does not depend on any of the parameters 
(masses, wave function renormalizations, couplings...) which appear in the action.
These cutoffs are logically at the opposite extreme of another class of cutoffs that depend on all
the couplings of the action and are called, for reasons that will be made clear
in section III, ``spectrally adjusted''.

In this note we would like to do two things.
The first is to calculate the properties of the gravitational
fixed point in the Einstein-Hilbert truncation using a pure cutoff, 
and compare them to the results of other, more popular cutoffs.
This we shall do in section IV.
The second is to show in general that whenever one uses a pure cutoff, 
there is a choice of parameters
that will yield the same FP as a spectrally adjusted cutoff.
This will be the subject of sect. V.

For definiteness we shall discuss scheme-dependence in the context of
Wetterich's exact RG equation \cite{Wetterich}, which uses an additive infrared cutoff.
To limit the dimension of the parameter space that we explore
we shall keep other aspects of the cutoff and gauge choice fixed.
Although stimulated by research on gravity, 
the points we make in sect. V are quite general
and should apply also to other quantum field theories 
and to other flow equations.

\section{The flow equation}

As mentioned in the Introduction, most of the progress towards asymptotic safety
of the last ten years has come from applying functional renormalization group methods to gravity.
In this section we describe these tools, using first the example of a real scalar field
and then describing some of the modifications that are needed in the application to gravity.
A general idea put forward by Wilson is that the functional integration should not be
performed in one single step covering all field fluctuations from the UV to the IR,
weighting all fluctuations with the same bare action,
but rather in a sequence of finite steps, updating the action at each step.
A concrete implementation of this idea that is easily amenable to explicit
calculations was given in 1993 by Wetterich \cite{Wetterich}.
We begin from a formal functional integral
\begin{equation}
\label{eq:Wk_def}
e^{-W_k[j]}=\int (d\phi) e^{- \left( S(\phi)+\Delta S_k(\phi)+\int j\phi \right) } \, \ ,
\end{equation}
where $j$ is an external source and
\begin{equation}
\label{cutoff}
\Delta S_k(\phi)=\frac{1}{2}\int d^4q \phi(-q) R_k(q^2) \phi(q)
\end{equation}
The effect of the new term $\Delta S_k$ is simply to modify the (inverse)
propagator of the theory: it replaces $q^2$ by 
\begin{equation}
\label{pk}
P_k(q^2)=q^2+R_k(q^2)
\end{equation}
The kernel $R_k(q^2)$ is chosen so as to suppress the propagation
of the modes with momenta $|q|\ll k^2$ and tends to zero for $|q|\gg k^2$
so that high momentum modes are integrated out without any suppression.
We shall discuss the properties of this function in greater detail in section III.
One then defines a scale--dependent effective action functional $\Gamma_k(\phi)$,
as the Legendre transform of $W_k$,
minus the term $\Delta S_k$ that we introduced in the beginning:
\begin{equation}
\label{Gamma}
\Gamma_k[\phi]=W_k[j]-\int j\phi-\Delta S_k(\phi) \, \ ,
\end{equation}
where $\phi$ is now to be interpreted as a shorthand for $\langle\phi\rangle$,
the variable conjugated to $j$.
If the functional integral is defined by an UV cutoff, then when $k$ tends to
this cutoff the average effective action is related by a nontrivial transformation
to the bare action \cite{elisa1}.
For $k\to 0$, $\Delta S_k\to 0$ and one
recovers the standard definition of the effective action (the generating function
of one--particle--irreducible Green functions).
It is not exactly the Wilsonian action but its definition is similar in spirit
and it is somewhat easier to work with.
If one evaluates this functional at one loop, it is
\begin{equation}
\label{eq:Gamma_1loop}
\Gamma_k^{(1)}=\frac{1}{2}\STr \log\left(\frac{\delta^2 S}{\delta\phi\delta\phi}+R_k\right)
\end{equation}
and its scale dependence is given by
\begin{equation}
\label{eq:Gamma_1loop_tder}
k\frac{d\Gamma_k}{dk}=
\frac{1}{2}
\STr\left({\delta^2 S\over\delta\phi\delta\phi}+R_k\right)^{-1}
k\frac{dR_k}{dk}\ .
\end{equation}
Here STr is a trace
that includes a factor $-1$ for fermionic fields and a factor $2$ for
complex fields.
It can be shown that the ``renormalization group improvement'' of this equation,
which consists in replacing $S$ by $\Gamma_k$ in the r.h.s.,
leads actually to an exact equation often called the Functional Renormalization Group Equation
(FRGE) \cite{Wetterich}:
\begin{equation}
\label{eq:FRGE}
k\frac{d\Gamma_k}{dk}=
{1\over 2}
\STr\left({\delta^2 \Gamma_k\over\delta\phi\delta\phi}+R_k\right)^{-1}
k\frac{dR_k}{dk}\ .
\end{equation}

It is important to observe that the last term in eq.(\ref{eq:FRGE}) suppresses the contribution
of high momentum modes so that the trace is ultraviolet finite: there is no need
to use any ultraviolet regularization.
In fact, once the equation has been derived, it is actually not necessary to
refer to the functional integral anymore.
The FRGE defines a flow in the space of all theories and if we start from any
point and we follow the flow in the limit $k\to 0$, then we find the
effective action, from which in principle we can derive everything we may
want to know about the theory.
Conversely, by following the flow towards higher energy we can establish
whether the theory has a FP with the desired properties.

The application of this equation to gravity has been discussed first in \cite{Reuter}.
Since gravity is a gauge theory, one has to take into account the complications
due to the gauge fixing and ghost terms.
So far the best way to deal with these complications is to use the background field method.
Let $\bar g_{\mu\nu}$ be a fixed but otherwise arbitrary metric. We can write
$g_{\mu\nu}=\bar g_{\mu\nu}+h_{\mu\nu}$. It is not implied that $h$ is small.
We choose a gauge--fixing condition
\begin{equation}
\label{eq:GF_act}
S_{GF}(\bar g, h)= \frac{1}{2} \int d^4x \, \sqrt{\bar g}\,\chi_{\mu}Y^{\mu\nu}\chi_{\nu}
\end{equation}
where $\chi_{\nu}=\nabla^{\mu}h_{\mu\nu}+\beta\nabla_{\nu}h$, $h= h_{\mu}^{\mu}$
and $Y$ is some operator, which in the simplest cases is just equal to $\bar g_{\mu\nu}$.
The standard formal manipulations in the path integral give rise to a ghost term
\begin{equation}
\label{eq:GH_act}
S_c=\int d^4x \, \sqrt{\bar g}\,\bar{C}_{\nu}(\Delta_{gh})_{\mu}^{\nu}C^{\mu}\ ,
\end{equation}
and, if $Y$ contains derivatives, also a ``third ghost'' term \cite{Buchbinder, CPR2}
\begin{equation}
\label{eq:3rdghost}
S_b=\frac{1}{2}\int d^4x \, \sqrt{\bar g}\,b_{\mu}Y^{\mu\nu}b_{\nu}\ .
\end{equation}
Also the cutoff term $\Delta S_k$ is written in terms of the background metric
\begin{equation}
\label{eq:cutoff_grav}
\Delta S_k(\bar g)=\int d^4x \, \sqrt{\bar g} \,
h_{\mu\nu}\bar g^{\mu\rho}\bar g^{\nu\sigma}R_k(\bar\Delta)h_{\rho\sigma}
\end{equation}
where $\bar\Delta$ is some differential operator constructed with the
background metric.
In this way one constructs a generating functional $W(j^{\mu\nu},\bar g_{\mu\nu})$
depending on sources that couple linearly to $h_{\mu\nu}$, and on the   
background metric.
Applying the definition eq. (\ref{Gamma}) one obtains a functional $\Gamma_k(h_{\mu\nu},\bar g_{\mu\nu})$
where $h_{\mu \nu}$ is now a shorthand for $\langle h_{\mu \nu}\rangle$, the Legendre conjugate of $j^{\mu\nu}$.
One can also think of $\Gamma_k$ as a functional of two metrics, namely
$\langle g_{\mu\nu}\rangle=\bar g_{\mu\nu}+\langle h_{\mu\nu}\rangle$
and the background metric.
In the limit $k\to 0$ this functional becomes the ordinary gravitational
effective action in the background gauge.
The functional $\Gamma_k(g,\bar g)$ is invariant under simultaneous
coordinate transformations of $g$ and $\bar g$, the so--called background
gauge transformations.
We will restrict our attention to the functional
$\Gamma_k(g)=\Gamma_k(g,g)$ obtained by the identification of the background field
(which hitherto remained completely unspecified) and the vacuum expectation value $g$.
By construction this functional has the same gauge invariance as the
original action and it contains the information about the familiar terms such
as the Einstein--Hilbert action. The functional $\Gamma_k(g,\bar g)$ 
contains in addition the information about the $k$--dependence of the gauge--fixing terms
and other genuinely bimetric terms in the action \cite{elisa2}.
In the following we will ignore the RG flow of these terms.
The functional $\Gamma_k(g,\bar g)$ obeys an FRGE that has the same form
as in eq. (\ref{eq:FRGE}), where $\phi$ now stands for $h_{\mu\nu}$, $C_{\mu}$, $\bar C_{\mu}$ and $b$.

\section{Cutoff schemes}

There is a large arbitrarines in the choice of the cutoff $\Delta S_k$.
Let us begin by discussing the simplest example:
a scalar theory in flat space with the action truncated to:
\begin{equation}
\label{eq:scalar_act}
S=\int dx\left(\frac{1}{2} \left( \partial \phi \right) ^2+V( \phi^2) \right)
\end{equation}
where $V$ is an even potential.
The cutoff is usually written in momentum space as in eq. (\ref{cutoff})
where $R_k$ is a function that is only required to satisfy the boundary
conditions that are appropriate to a cutoff, namely:
\begin{itemize}
\item it tends rapidly to zero when $q^2>k^2$.
\item it tends to $k^2$ when $q^2$ tends to zero.
\item it is monotonically decreasing in $q^2$.
\item it is monotonically decreasing in $k$, uniformly in $q^2$.
\end{itemize}
The second condition is a kind of normalization and is not strictly necessary:
it is possible to allow $R_k$ to tend to some other finite value, or even
to infinity, when $k\to 0$. However, for reasons that will become clear shortly,
it is convenient to keep fixed the value of $R_k(0)$.
Aside from this, the shape of the function is arbitrary.
Inserting the ansatz \eqref{eq:scalar_act} for $\Gamma_k$, the FRGE \eqref{eq:FRGE} becomes
\begin{equation}
\label{eq:FRGE_scalar}
k\frac{d\Gamma_k}{dk}=\frac{1}{2}\frac{1}{(4\pi)^2}
\int dz z \, \frac{\partial_t R_k(z)}{P_k(z)+ 2V'+4\phi^2 V''} 
\, \ ,
\end{equation}
where $z=q^2$, $t=\log(k/k_0)$ 
and the modified inverse propagator $P_k$ is defined as in eq. (\ref{cutoff}).

From here, taking derivatives with respect to $\phi^2$, one can
extract the beta functions of the coefficients of the Taylor expansion
of the potential.
These beta functions will depend in general on the choice of $R_k$.
One notable exception, in four dimensions, is the beta function of the
quartic coupling, which in the limit where the mass and the higher couplings are negligible,
turns out to be independent of $R_k$.
The way in which this happens is as follows: the trace in the r.h.s.
of the FRGE involves an integration over $q^2$.
In the specific case, the function to be integrated turns out to be a total
derivative and therefore depends only on the boundary values.
Since these have been fixed, it is completely determined.
See Appendix A in \cite{CPR2} for details.

The discussion can be refined by taking into account a wave function renormalization
constant $Z$ in front of the kinetic term.
One then usually modifies the definition eq. (\ref{cutoff}) by including a factor $Z$:
\begin{equation}
\label{eq:scalar_cutoffZ}
\Delta S_k= \frac{Z}{2} \int  dq \, \phi(-q) \, R_k(q^2) \, \phi(q)\ .
\end{equation}
The reason why this is convenient is that the inverse propagator $Z q^2$ and
the cutoff $ZR_k(q^2)$ then neatly combine into a modified inverse propagator $Z P_k(q^2)$
and the overall factors of $Z$ cancel between numerator and denominator of the FRGE,
which takes the form
\begin{equation}
\label{eq:FRGE_scalarZ}
k\frac{d\Gamma_k}{dk}=\frac{1}{2}\frac{1}{(4\pi)^2}
\int  dz z \, \frac{\partial_t R_k+\eta R_k(z)}{P_k(z)+\frac{2V'+4\phi^2V''}{Z}}
\end{equation}
where we have defined $\eta=\frac{d\log Z}{d\log k}$.

The choice of the shape of the function $R_k$ is an aspect of the cutoff scheme
which involves infinitely many parameters.
When one considers more complicated theories, further ``discrete'' choices present themselves.
For example, in the case of gravity truncated to the Hilbert action, the second variation of the action 
has the following  schematic form:
\begin{equation}
\label{schematic}
\Gamma^{(2)}_k =  a Z [-\Box + b R + c \Lambda ],
\end{equation}
where $Z=1/(16\pi G)$, $a$, $b$ and $c$ are constants and $\Box=\nabla^\mu\nabla_\mu$ 
and the curvature term $R$ are computed with the background metric. 
Tensor structures are omitted, since they are not essential for the argument.
Since in curved spacetime one cannot use momentum space methods, 
in the definition of the cutoff one has to decompose the fields on a basis of
eigenvectors of some selfadjoint differential operator.
The cutoff then suppresses the modes with eigenvalues $\lambda<k^2$
(if the operator is of second order). One then has a choice of what operator to use.
In \cite{CPR2} we called a cutoff of type I, II or III, respectively, 
if it is defined using the spectrum of $-\Box$, $-\Box+bR$ or $-\Box+bR+c\Lambda$.

In the literature, an overall factor of $Z=1/(16\pi G)$ is inserted in the definition of the cutoff as in 
eq. (\ref{eq:scalar_cutoffZ}).
In the case of gravity there are some differences.
Unlike in the scalar case, the coefficient $Z$ is dimensionful;
even more important, one can show that $Z$ is now an essential coupling constant
in the sense that it cannot be eliminated by rescaling the metric without 
changing the unit of mass $k$ \cite{Perini3}.
In spite of these differences, it is still true that $Z$ does not appear in the reference operator,
so in the case of type I and II cutoffs the spectrum of the operator does not depend on $k$.
In the type III cutoff the reference operator contains the running coupling $\Lambda$.
As a consequence, also the spectrum of the reference operator changes with $k$. 
For this reason this is called a ``spectrally adjusted cutoff''.
It has been argued in \cite{gies} that such cutoffs give improved results.

In this paper we would like to explore in the opposite direction and
consider a class of cutoff functions that do not depend on {\it any} coupling, not even on $Z$.
Quite generally, we will refer to a cutoff that does not contain any parameter that appears
in the action as a ``pure'' cutoff.

\section{Gravitational beta functions}

We will now discuss the example of gravity in the Einstein-Hilbert truncation,
where one retains only the terms with up to two derivatives of the metric:
\begin{equation}
\label{ehaction}
\Gamma_{k}=\int dx\sqrt{g}\,(2\Lambda Z-ZR)+S_{GF}+S_{\rm ghost}\ ,
\end{equation}
where $Z=1/(16\pi G)$, $S_{GF}$ is a gauge--fixing term and $S_{\rm ghost}$ is the ghost action.
We decompose the metric into $g_{\mu\nu}=g_{\mu\nu}^{(B)}+h_{\mu\nu}$
where $g_{\mu\nu}^{(B)}$ is a background.
We consider a de Donder background gauge:
\begin{equation}
\label{eq:EH_GF}
S_{GF}(g^{(B)},h)=\frac{Z}{2\alpha}\int dx\sqrt{g^{(B)}}\chi_{\mu}g^{(B)\mu\nu}\chi_{\nu}\ ,
\end{equation}
where
$$
\chi_{\nu}=\nabla^{\mu}h_{\mu\nu}-\frac{1}{2}\nabla_{\nu}h\ .
$$
All covariant derivatives are with respect to the background metric.
In the following all metrics will be background metrics, and we will
omit the superscript $(B)$ for notational simplicity.
In order to restrict the number of parameters, we will fix $\alpha=1$
which leads to considerable simplifications.

After calculating the second variation of the action, we decompose $h_{\mu\nu}$
into its different spin components according to
\begin{equation}
\label{decomposition}
h_{\mu\nu}=h^T_{\mu\nu}+\nabla_{\mu}\xi_{\nu}
+\nabla_{\nu}\xi_{\mu}+\nabla_{\mu}\nabla_{\nu}\sigma-\frac{1}{
d}g_{\mu\nu}\Box\sigma+\frac{1}{d}g_{\mu\nu}h. 
\end{equation}
and
\begin{equation}
\label{ghostdecomposition}
C^{\mu}=C^T{}^{\mu}+\nabla^{\mu} C\ \ , \qquad
{\bar C}_{\mu}={\bar C}^T_{\mu}+\nabla_{\mu} {\bar C}\ ,
\end{equation}
where $h^T_{\mu\nu}$ is tranverse and traceless, $\xi$ is a transverse vector,
$\sigma$ and $h$ are scalars, $C^T$ and $\bar C^T$ are transverse vectors, and $C$ and $\bar C$ 
are scalars.
These fields are subject to the following differential constraints:
\begin{equation*}
h_{\mu}^{T\mu}=0\ ; \qquad \nabla^{\nu}h_{\mu\nu}^{T}=0\ ;
\qquad \nabla^{\nu}\xi_{\nu}=0\ ; \qquad \nabla^\mu \bar C^{T}_{\mu}=0\ ; \qquad
\nabla_\mu C^{T\mu}=0\ .
\end{equation*}
We further redefine 
\begin{eqnarray}
\label{redefinitions}
\xi_{\mu}&=&\sqrt{-\Box-\frac{R}{d}}\,\hat\xi_{\mu} \, \ , \qquad
\sigma=\sqrt{-\Box}\sqrt{-\Box-\frac{R}{d-1}}\hat\sigma.
\end{eqnarray}
which removes some powers of $-\Box$ from the second variation
and furthermore cancels some Jacobian determinants that arise
in the functional integral when one performs the decomposition eq. (\ref{decomposition}).
We then drop the hat from $\xi$ and $\sigma$

In order to extract the beta functions of $\Lambda$ and $Z$
it is enough to calculate the r.h.s. of the FRGE on the background of
a sphere. Then the inverse propagator diagonalizes
\begin{eqnarray}
\label{kinIb}
\Gamma^{(2)}_{h^T_{\mu\nu}h^T_{\rho\sigma}}&=&
\frac{Z}{2}\left[-\Box+\frac{2}{3}R-2\Lambda\right]
\frac{1}{2} ( g^{\mu \rho} \, g^{\nu \sigma} + g^{\mu \sigma} \, g^{\nu \rho} )
\nonumber\\
\Gamma^{(2)}_{\xi_{\mu}\xi_{\nu}}&=&
Z\left[-\Box+\frac{1}{4}R-2\Lambda\right]g^{\mu\nu}\nonumber\\
\Gamma^{(2)}_{hh}&=&
-\frac{Z}{8}\left[-\Box-2\Lambda\right]\nonumber\\
\Gamma^{(2)}_{\sigma\sigma}&=&
\frac{3 Z}{8}\left[-\Box-2\Lambda\right]\nonumber\\
\Gamma^{(2)}_{{\bar C}^T_{\mu}{C}^T_{\nu}}&=&
\left[\Box+\frac{1}{4}R\right]g^{\mu\nu}\nonumber\\
\Gamma^{(2)}_{{\bar C}{C}}&=&
\left[\Box+\frac{1}{2}R\right]
\end{eqnarray}
Now we define the cutoff. We require that after adding the cutoff,
the modified inverse propagator has the same form as in \eqref{kinIb}
except for the replacement of $-\Box$ by $P_k(-\Box)$,
where $P_k$ is defined as in eq. (\ref{pk}).
This means that when the inverse propagator has the form eq. (\ref{schematic}),
the cutoff is
\begin{equation}
\label{cutoffIb}
\calr_k=aZ R_k(-\Box)\ .
\end{equation}

This type of cutoff was introduced in \cite{Dou} and studied in greater
generality in \cite{Lauscher}.
With this cutoff the r.h.s. of the FRGE takes the form
\begin{eqnarray}
\label{withred}
 \frac{d \Gamma_k}{dt}
&=& \frac{1}{2} \textrm{Tr}_{(2)} \frac{\partial_t R_k+\eta
R_k}{P_k+\frac{2}{3}R-2\Lambda}
+\frac{1}{2} \textrm{Tr}'_{(1)}
\frac{\partial_t R_k+\eta R_k}{P_k+\frac{1}{4}R-2\Lambda}\nonumber\\
&&+\frac{1}{2} \textrm{Tr}_{(0)} 
\frac{\partial_t R_k+\eta R_k}{P_k-2\Lambda}
+\frac{1}{2} \textrm{Tr}''_{(0)} 
\frac{\partial_t R_k+\eta R_k}{P_k-2\Lambda}
%
-\textrm{Tr}_{(1)} \frac{\partial_t R_k}{P_k-\frac{R}{4}}
-\textrm{Tr}'_{(0)} \frac{\partial_t R_k}{ P_k- \frac{R}{2}}\ .
\end{eqnarray}
where $\eta=\frac{d\log Z}{d\log k}$.
The first term comes from the spin--2, transverse traceless
components, the second from the spin--1 transverse vector, the third
and fourth from the scalars $h$ and  $\sigma$.
The last two contributions come from the transverse and
longitudinal components of the ghosts. A prime or a double prime
indicate that the first or the first and second eigenvalues have to
be omitted from the trace. The reason for this is explained in
the references quoted above.

In order to be able to perform the calculation
in closed form we choose the optimized cutoff $R_k(z)=(k^2-z)\theta(k^2-z)$ \cite{optimized}.
In this way one finds the following beta functions for the dimensionless
couplings $\tilde Z=Z/k^2$, $\tilde\Lambda=\Lambda/k^2$:
\begin{eqnarray}
\label{betasIbwfr}
\partial_t \tilde Z
&=&-2\tilde{Z} + \frac{373 -654 \tilde{\Lambda} + 600 \tilde{\Lambda}^2 }{1152 \pi^2 (1 - 2 \tilde{\Lambda} )^2}
+ \frac{\partial_t \tilde{Z}}{\tilde{Z} } \frac{29- 9 \tilde{\Lambda} }{1152 \pi^2 (1 - 2 \tilde{\Lambda} )^2}
\nonumber\\
\partial_t (\tilde{Z} \tilde{\Lambda} )
&=& -4 \tilde{Z} \tilde{\Lambda} + \frac{1 + 3 \tilde{\Lambda} }{12 \pi^2 (1 - 2 \tilde{\Lambda} )}
+ \frac{\partial_t \tilde{Z}}{\tilde{Z} } \frac{5}{ 192 \pi^2 (1 - 2 \tilde{\Lambda} )}
\ .
\end{eqnarray}
which admit a FP at $\tilde G_*=0.7012$, $\tilde\Lambda_*=0.1715$, with critical exponents
$\theta=1.689\pm 2.486 i$.

\begin{figure}
[t]\center
{
\resizebox{0.45 \columnwidth}{!}
{\includegraphics{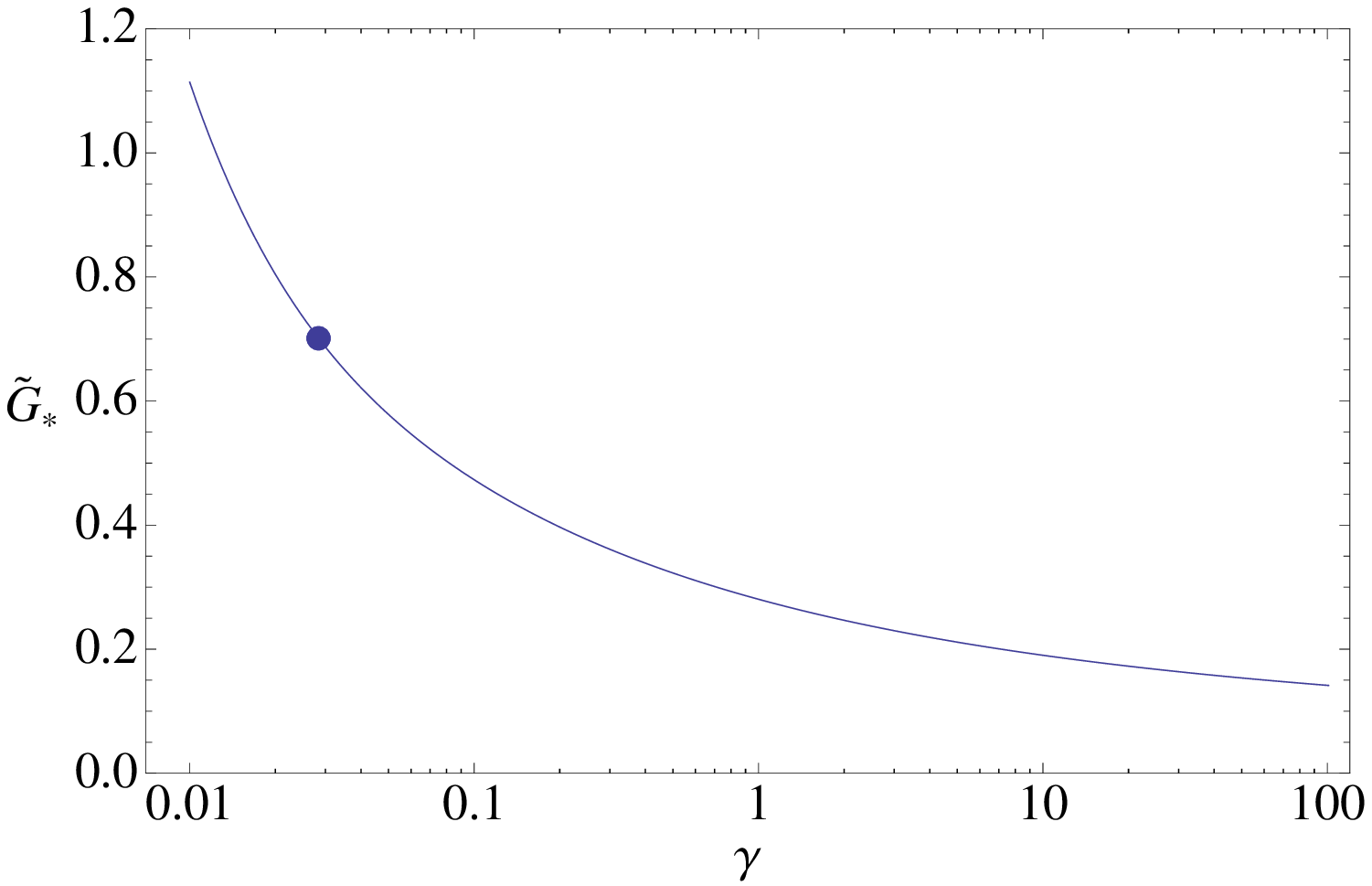}}
\resizebox{0.45 \columnwidth}{!}
{\includegraphics{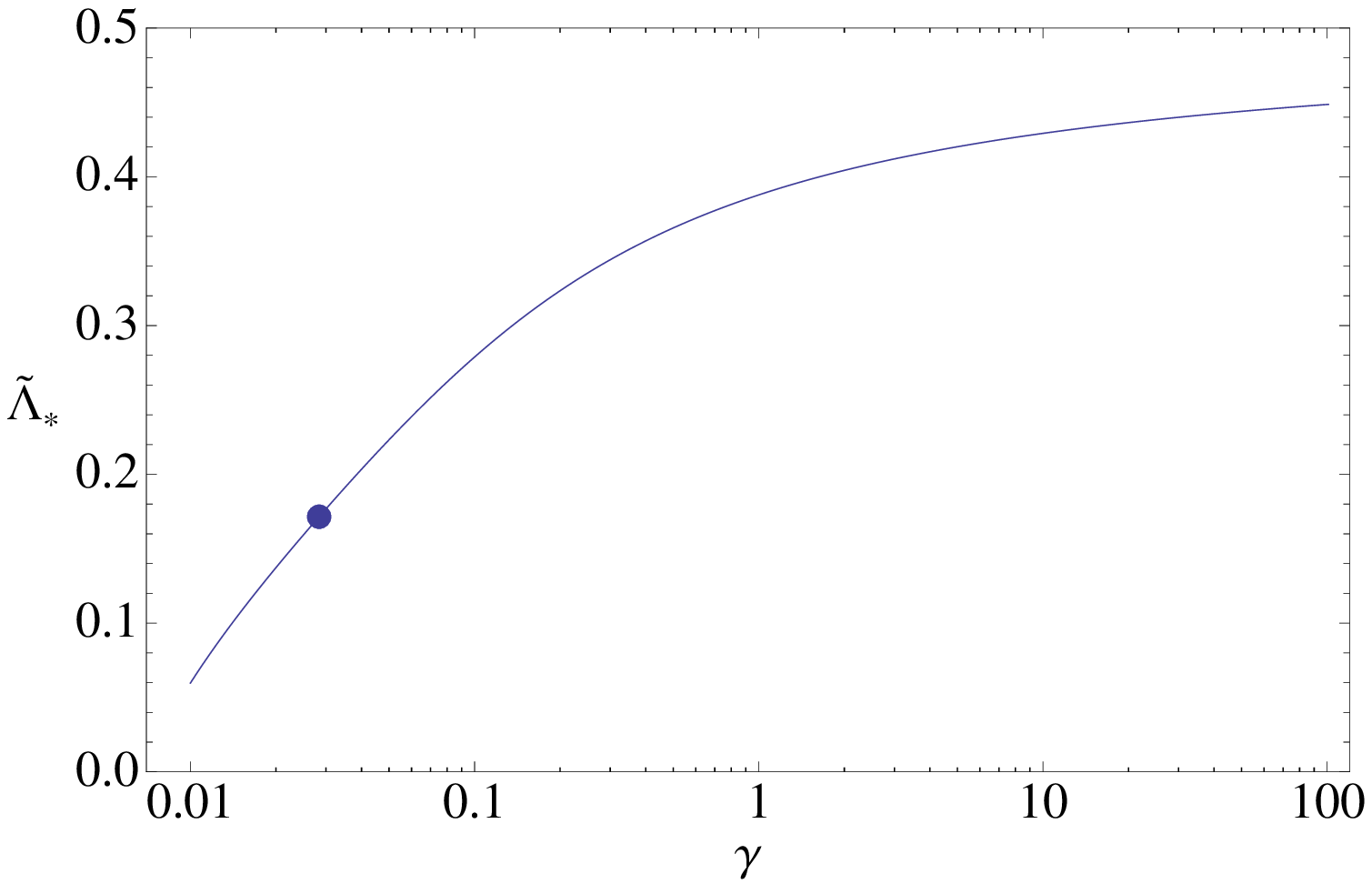}}
\resizebox{0.6 \columnwidth}{!}
{\includegraphics{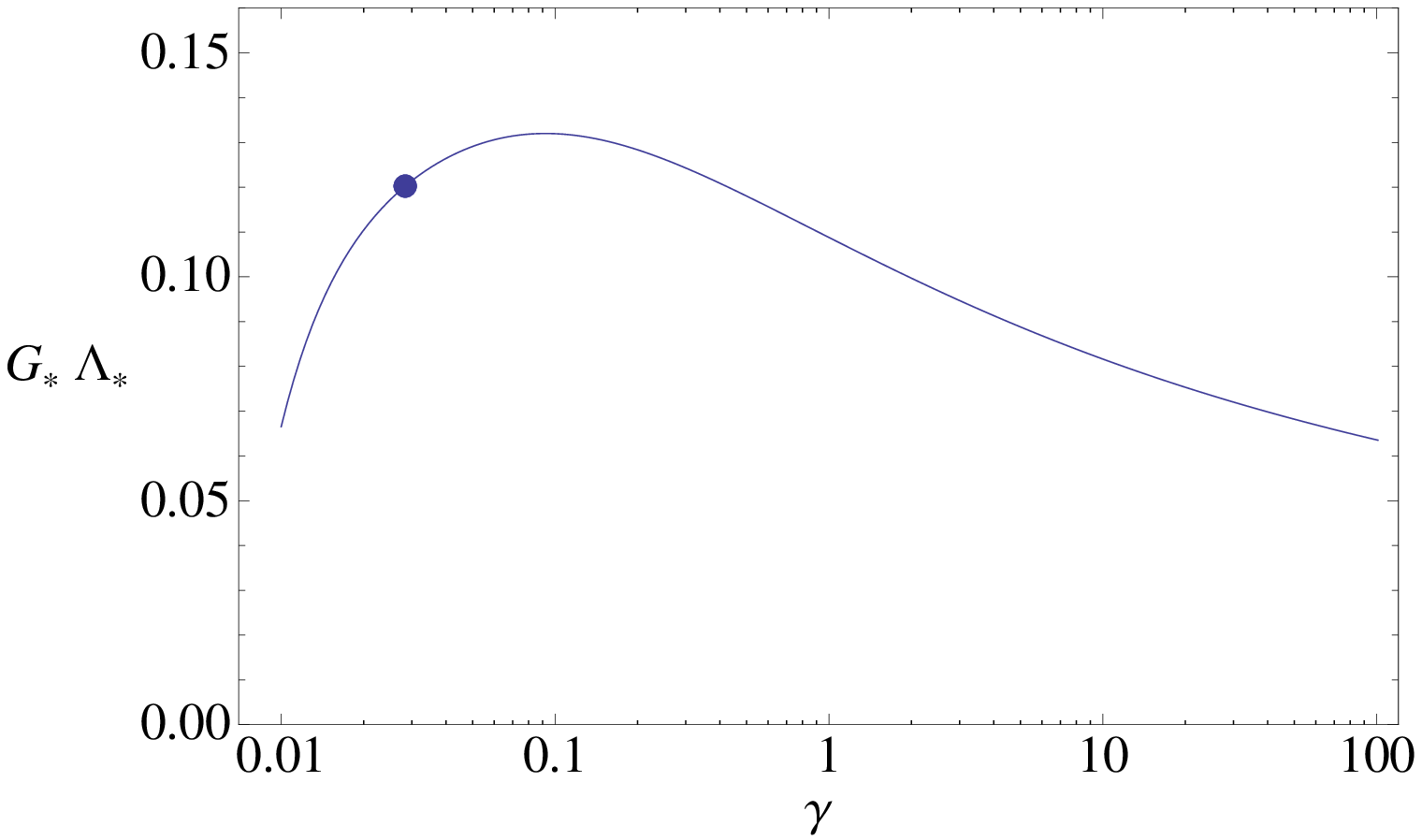}}
\caption{\label{fig:1}Value of $\tilde G_*$ (left panel), $\tilde\Lambda_*$ (right panel)
and  $\Lambda_* G_*$ (lower panel) as functions of $\gamma$ with   
a pure cutoff. The dot indicates the values for the type Ib cutoff.}
}
\end{figure}

As anticipated, we would like to examine a different type of cutoff,
not depending on any of the parameters that are present in the action.
The cutoff eq. (\ref{cutoffIb}) depends on the parameter $Z$,
so to define a pure cutoff we replace $Z$ by $\gamma k^2$
where $\gamma$ is an arbitrary number:
\begin{equation}
\label{cutoffpure}
\calr_k=a \gamma k^2 R_k(-\Box)\ .
\end{equation}
The FRGE now reads
\begin{eqnarray}
\label{withred}
 \frac{d \Gamma_k}{dt}
&=& \frac{1}{2} \textrm{Tr}_{(2)}
\frac{\partial_t R_k+2R_k}{\frac{Z}{\gamma k^2}\left(-\Box+\frac{2}{3}R-2\Lambda\right)+R_k(-\Box)}
+\frac{1}{2} \textrm{Tr}'_{(1)}
\frac{\partial_t R_k+2R_k}{\frac{Z}{\gamma k^2}\left(-\Box+\frac{1}{4}R-2\Lambda\right)+R_k(-\Box)}\nonumber\\
&&+\frac{1}{2} \textrm{Tr}_{(0)} 
\frac{\partial_t R_k+2 R_k}{\frac{Z}{\gamma k^2}\left(-\Box-2\Lambda\right)+R_k(-\Box)}
+\frac{1}{2} \textrm{Tr}''_{(0)} 
\frac{\partial_t R_k+2 R_k}{\frac{Z}{\gamma k^2}\left(-\Box-2\Lambda\right)+R_k(-\Box)}\nonumber\\
&&-\textrm{Tr}_{(1)} \frac{\partial_t R_k}{P_k-\frac{R}{4}}
-\textrm{Tr}'_{(0)} \frac{\partial_t R_k}{ P_k- \frac{R}{2}}\ .
\end{eqnarray}

\begin{figure}
[t]\center
{
\resizebox{0.6 \columnwidth}{!}
{\includegraphics{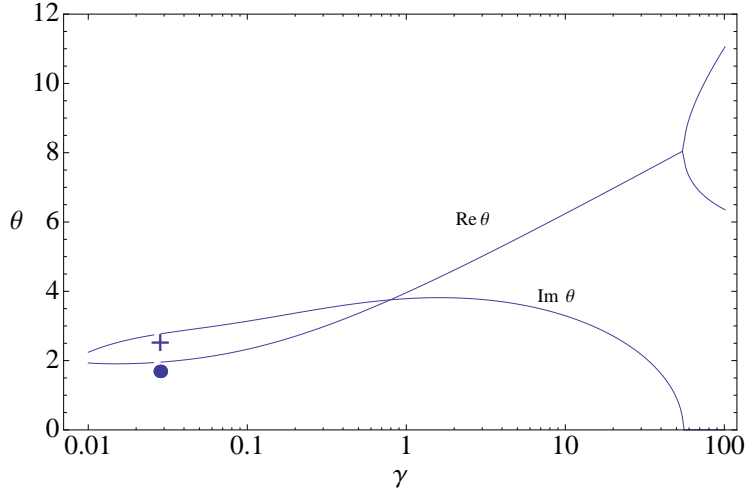}}
\caption{\label{fig:2}
Real and imaginary parts of the critical exponents as functions of $\gamma$.
The dot and the cross indicate the real and imaginary part of the critical exponent 
for the type Ib cutoff.}
}
\end{figure}

This leads to the following beta functions
\begin{eqnarray}
\label{betasIbwfr}
\partial_t \tilde Z
&=&-2\tilde{Z} + \frac{49 \gamma (\gamma - \tilde{Z}) + ( 1 -2 \tilde{\Lambda} ) 
(25 \tilde{Z}^2 - 151 \tilde{Z} \gamma + 28 \gamma^2)}{192\pi^2(\gamma-\tilde{Z})^2(1-2\tilde{\Lambda})} 
\nonumber \\
&&
\qquad\qquad\qquad\qquad 
-\frac{\gamma\left[3(\gamma-\tilde{Z})^2+\tilde{Z}(1-2\tilde{\Lambda})(101\tilde{Z}-3\gamma)\right]}
{192 \pi^2 (\gamma - \tilde{Z})^3}
\, X
\nonumber\\
\partial_t (\tilde{Z} \tilde{\Lambda} )
&=& -4 \tilde{Z} \tilde{\Lambda}-
\frac{9\gamma^2+4\tilde{Z}^2-\gamma \tilde{Z}(23-20\tilde{\Lambda} )}{32 \pi^2 (\gamma-\tilde{Z})^2 }
-\frac{5\gamma\left[\gamma^2-2\gamma\tilde{Z}+4\tilde{Z}^2\tilde{\Lambda}(1-\tilde{\Lambda})\right]}
{16 \pi^2(\gamma-\tilde{Z})^3}\, X
\ .
\end{eqnarray}
where $$ X = \ln \left( \frac{\tilde{Z} (1- 2 \tilde{\Lambda} ) }{\gamma - 2 \tilde{Z} \tilde{\Lambda} } \right).$$
The appearance of the logarithms is due to the mismatch between the coefficients
of $-\Box$ and $R_k(-\Box)$, which leaves some explicit terms with $z=-\Box$ to be integrated over. 
The FP now depends on the arbitrary parameter $\gamma$, which is part of the
freedom in the definition of the cutoff.
This reflects itself in the position of the fixed point, as shown in Figure 1.
We give separately the dependence of $\tilde G_*$, $\tilde\Lambda_*$ and of
the dimensionless product $\Lambda_* G_*$.
We see that as $\gamma$ varies over four orders of magnitude,
$\tilde G_*$, $\tilde\Lambda_*$ each vary by less than one order of magnitude,
and $\Lambda_* G_*$ changes just by a factor smaller than 2.
It had been observed before that the dimensionless product $\Lambda G$
has a beta function that is gauge independent in lowest order
in an expansion in $\tilde\Lambda$ \cite{Dou};
also its value at the FP was found to be
quite insensitive to the choice of gauge and cutoff.
Our findings confirm this picture also for the dependence on the parameter $\gamma$.
In figure 2 we also give the critical exponents as functions of $\gamma$.
As with other cutoffs, they form a complex conjugate pair,
but for large $\gamma$ the imaginary part of the eigenvalue goes to zero
and for $\gamma>60$ we find two real eigenvalues.
Clearly for very large or very small $\gamma$ the properties of the FP are
significantly affected, but there is a wide range of values for which
the properties of the FP are quite stable.

We observe that the curves in Figure 1 pass through the position of the
fixed point in the type Ib cutoff examined previously,
which is marked by a dot in the graphs.
In other words, there is a value of $\gamma$ for which the
pure cutoff gives a FP in the same position as the type Ib cutoff.
The corresponding value is precisely $\gamma=\tilde Z_*=0.0284$.
It would seem from (28) that the beta functions become singular
when $\gamma=\tilde Z$ but if we put $\gamma=\tilde Z+\epsilon$
and expand in powers of $\epsilon$, the coefficient of the negative
powers of $\epsilon$ cancel out.
Furthermore, one finds that the leading ($\epsilon$-independent) terms
in the expansion coincide with the first two terms on the r.h.s. of (25).
There is clearly something special happening when $\gamma$ has the value
of $\tilde Z_*$ in the type Ib cutoff.
In the next section we will explain the origin of this coincidence 
and we will see that it holds in much greater generality.

\section{Fixed points with pure and spectrally adjusted cutoffs}

We have shown in the previous section, in the special case of gravity in the
Einstein Hilbert truncation, that there is a value of the parameter $\gamma$ for which
a pure cutoff reproduces the results of a type I cutoff.
In this section we argue that this property is quite general.
Since we are going to move to a much more general setting we will have to
make some choices in order to somehow circumscribe the scope of the argument.

We assume that $\Gamma_k$ admits a derivative expansion of the form
\begin{equation}
\label{expansion}
\Gamma _k=  \sum_i g_i \mathcal{O}_i,
\end{equation}
where $\calo_i$ are operators and $g_i$ are numerical parameters depending on $k$.
The operators $\calo_i$ are integrals of the form
\begin{equation}
\label{eq:funcO_i}
\mathcal{O}_i=\int\,{d}^{d}x\sqrt{g}\Omega_i,
\end{equation}
where $\Omega_i$ are (possibly nonpolynomial) functions of the fields and their derivatives,
respecting all the symmetries that the theory is supposed to possess.
In gauge theories, $\Omega_i$ are constructed with covariant derivatives and curvatures.
The number of derivatives increases with $i$, but the precise correspondence need not
be spelled out here.

Some of the parameters appearing in the expansion may be eliminated by field redefinitions.
This is the case, for example, for the wave function renormalization constants.
Such parameters are said to be ``redundant'' or ``inessential'' \cite{wegner}.
We assume that the theory has been parametrized in such a way that a certain
subset of the $g_i$ is redundant, while the remaining ones are ``essential''.

The fields do not have any scale dependence, so that
\begin{equation}
\label{eq:betacoup}
k\frac{d\Gamma _k}{dk}=  \sum_i \beta_i \mathcal{O}_i,
\end{equation}
where $\beta_i(g_j,k)=k\frac{dg_i}{dk}$
are the beta functions. In general they depend on all the $g_i$ and also explicitly on $k$.
Note that we call ``beta functions'' the derivatives of the parameters appearing
in the action whether they are essential or not.
One sometimes prefers to call ``anomalous dimensions'' the (logarithmic) derivatives of irrelevant 
parameters such as the wave function renormalization constants.
We will not need to make this terminological distinction here.
We will call $\beta$ (without subscript $i$) the ``beta functional'' on the r.h.s. of the FRGE
$$
\beta= \sum_i \beta_i \mathcal{O}_i\ .
$$

If the operator $\Omega_i$ has dimension $\alpha_i$, $\calo_i$
has dimension $\alpha_i-d$ and $g_i$ has dimension $d_i=d-\alpha_i$.
One can now define dimensionless couplings $\tilde g_i$ and 
dimensionless operators $\tilde\calo_i$ by $g_i=k^{d_i}\tilde{g}_i$ and
$\calo_i = k^{-d_i}\tilde{\calo_i}$,
so that eq. (\ref{expansion}) can also be written as $\Gamma_k=\sum_i \tilde g_i \tilde\calo_i$.
The condition that has to be satisfied by a FP is
\begin{equation}
\label{fpeqi}
k\frac{d\tilde g_i}{dk}=0\ ,
\end{equation}
for all essential couplings $g_i$.
We can rewrite this as follows.
From the definition of $\tilde g_i$ we obtain
$\partial_t g_i=d_i g_i+k^{d_i}\partial_t\tilde g_i$.
Then we can rewrite eq. (\ref{eq:betacoup}) as
$$
k\frac{d\Gamma _k}{dk}=\sum_i d_i\tilde g_i\tilde\calo_i+\sum_i \partial_t\tilde g_i\tilde\calo_i\ .
$$
Then the FP equation can be written compactly as
\begin{equation}
\label{fpeq}
\left(-\sum_i d_i\tilde g_i\tilde\calo_i+\beta\right)\Bigg|_{\mathrm{essential}}=0
\end{equation}
where the subscript ``essential'' means that the equation has to be projected
on the subspace of essential couplings.
The individual equations eq. (\ref{fpeqi}) can be obtained from the functional
equation by extracting the coefficient of the operator $\tilde\calo_i$.
We will now compare the functional form of this equation for two classes of cutoffs.

For definiteness we start by choosing a ``spectrally adjusted'' cutoff, 
defined as follows.
The second variation of the action is a differential operator
\begin{equation}
\label{delta}
\Delta(g_i)=\frac{\delta^2\Gamma_k}{\delta\phi\delta\phi}
=\sum_i g_i \frac{\delta^2\calo_i}{\delta\phi\delta\phi}\ .
\end{equation}
By this notation we emphasize that the operator depends on all the parameters $g_i$.
In the case of gauge theories the operator $\Delta$ is constructed with the covariant derivative $\nabla_\mu$.
We choose the cutoff $\calr$ to be a function of the full operator $\Delta$:
$\calr_k=R_k(\Delta(g_i))$,
where $R_k$ is one of the functions that were discussed in section III.
Then the modified inverse propagator is
\begin{equation}
\label{gencutoff}
\Delta(g_i)+R_k(\Delta(g_i))=P_k(\Delta(g_i))\ ,
\end{equation}
where $P_k$ is defined as in eq. (\ref{pk}).
This is a cutoff of ``type III'', in the terminology of \cite{CPR2}.
It is spectrally adjusted because it manifestly depends on all
the couplings, masses and wave function renormalizations.

The r.h.s. of the FRGE can now be written
\begin{equation}
\label{rhs1}
\beta=\frac{1}{2}\STr
\biggl(\Delta(g_i)+R_k(\Delta(g_i))\biggr)^{-1}
\left(\frac{\partial R_k(\Delta(g_i))}{\partial t}
+R'_k(\Delta(g_i))\frac{\partial\Delta}{\partial g_i}\beta_i\right)\ ,
\end{equation}
where a prime indicates the derivative of the function with respect to its argument.
In the first term one derives only the explicit dependence of the cutoff on $k$
and in the second the dependence that comes from the flow of the $g_i$.
From here the beta functions $\beta_i$ can be obtained in a two step procedure.
First one has to extract from eq. (\ref{rhs1}) the coefficient of $\calo_i$.
Formally we can write
$$
\beta_i=\frac{\delta\beta}{\delta \calo_i}
$$
This is usually the most labor-intensive part of the calculation,
but still it does not immediately give the beta function, because the r.h.s.
is itself a linear combination of the beta functions, of the form
$$
\frac{\delta\beta}{\delta \calo_i}=B_i+A_{ij}\beta_j\ .
$$
where $B_i$ are the one loop beta functions and $A_{ij}$ are calculable coefficients.
The beta functions can be obtained by solving this linear system:
$$
\beta_i=(\mathbf{1}-A)^{-1}_{ij}B_j\ .
$$
If one is only interested in the location of the FP,
one can avoid this step by the following trick \cite{Perini1}.
Since at a FP $g_i=\tilde g_{i*}k^{d_i}$, for some constants $\tilde g_{i*}$,
we obtain an equivalent set of FP equations if in the beta functional
we replace $\beta_i$ by $d_i g_i=d_i\tilde g_i k^{d_i}$.
This modified beta functional is
\begin{equation}
\label{rhs2}
\bar\beta=\frac{1}{2}\STr
\biggl(\Delta(g_i)+R_k(\Delta(g_i))\biggr)^{-1}
\left(
\frac{\partial R_k(\Delta(g_i))}{\partial t}
+R'_k(\Delta(g_i))\frac{\partial\Delta}{\partial g_i}d_i \tilde g_ik^{d_i}
\right)
\end{equation}
If we define
$$
\bar\beta_i=\frac{\delta\bar\beta}{\delta \calo_i}
$$
these expressions do not contain the $\beta$ functions anymore,
and so they can be plugged directly in the FP equation.
The FP equations obtained from the modified beta functions $\bar\beta_i$
have the same FP solutions as the ones obtained from the
true beta functions $\beta_i$.
We observe that the second term on the r.h.s. of eq. (\ref{rhs2}) is not just a function of $\Delta$.
In general it is a complicated operator that will not commute with $\Delta$ itself.
We actually do not have the mathematical tools to extract beta functions from
such complicated traces involving functions of several noncommuting operators.
However, calculability is not required here, so we can proceed formally.
We now simply assume that the FP equations determined in this way have a solution
at $\tilde g_i=\tilde g_{i*}$.

The source of $g_i$-dependence in the cutoff definition given above
is the operator $\Delta$.
We can turn the cutoff into a pure cutoff if we replace all the
couplings appearing in $\Delta$ by arbitrary constants,
multiplied by suitable powers of $k$ to preserve the correct dimensionalities.
The cutoff is then $R_k(\Delta(\gamma_ik^{d_i})))$.
With this cutoff the r.h.s. of the FRGE reads
\begin{equation}
\label{rhs3}
\beta=\frac{1}{2}\STr
\left(\Delta(g_i)+R_k(\Delta(\gamma_ik^{d_i}))\right)^{-1}
\left(\frac{\partial R_k(\Delta(\gamma_ik^{d_i}))}{\partial t}
+R'_k(\Delta(\gamma_ik^{d_i}))\frac{\partial\Delta}{\partial g_i}d_i\gamma_ik^{d_i}\right)
\end{equation}
From here one can extract beta functions
$$
\beta_i(g_i,\gamma_i)=\frac{\delta\beta}{\delta \calo_i}
$$
that can be used to write FP equations.
These FP equations depend parametrically on the arbitrary numbers $\gamma_i$.
Recalling that $g_i=\tilde g_i k^{d_i}$ and comparing eq. (\ref{rhs3})
to eq. (\ref{rhs2}) we see that the only difference lies in the replacement
of $\tilde g_i$ by $\gamma_i$ in certain functional dependences.

It is clear that since the FP equation for the spectrally adjusted cutoff
has a zero when we replace everywhere $\tilde g_i$ by the numbers $\tilde g_{i*}$,
then the FP equation for the pure cutoff will also have a zero when we replace
all the $\gamma_i$ and all the $\tilde g_i$ by $\tilde g_{i*}$.
Therefore with the particular choice of parameters $\gamma_i=\tilde g_{i*}$, 
the pure cutoff produces a FP in the same position as the spectrally adjusted 
type III cutoff.

This result has been derived using what we call a ``type III'' cutoff,
because the argument is easier to make independently of the form of
the action, but we believe that it holds more generally, also for other
cutoffs.
To illustrate this consider a generalization of what was called a ``type I''
cutoff in \cite{CPR2}.
In a gauge theory the second variation defined in eq. (\ref{delta}) is 
a differential operator constructed with the covariant derivative $\nabla_\mu$.
Let us assume that the truncation of the theory is such that $\Delta$
depends on $\nabla_\mu$ only through the combination $-\Box=-\nabla_\mu\nabla^\mu$.
To make this explicit let us write it as $\Delta(-\Box,g_i)$.
A generalized type I cutoff can be defined by the requirement that
the modified inverse propagator has the same form as the original one
except for the replacement of $-\Box$ by $P_k(-\Box)$:
\begin{equation}
\label{eq:gen_cutoff}
\calr_k(-\Box,g_i)=\Delta(P_k(-\Box),g_i)-\Delta(-\Box,g_i)\ .
\end{equation}
The beta functional that one obtains with this cutoff has the form
\begin{equation}
\label{rhs4}
\beta=\frac{1}{2}\STr
\left(\Delta(P_k(-\Box),g_i)\right)^{-1}
\left(\frac{\partial\Delta}{\partial(-\Box)}\frac{\partial P_k(-\Box)}{\partial t}
+\frac{\partial}{\partial g_i}
\biggl(\Delta(P_k(-\Box),g_i)-\Delta(-\Box,g_i)\biggr)\beta_i\right)
\end{equation}
which upon use of the trick explained above yields equivalent FP equations as
\begin{equation}
\label{rhs5}
\bar\beta=\frac{1}{2}\STr
\left(\Delta(P_k(-\Box),g_i)\right)^{-1}
\left(\frac{\partial\Delta}{\partial(-\Box)}\frac{\partial P_k(-\Box)}{\partial t}
+\frac{\partial}{\partial g_i}
\biggl(\Delta(P_k(-\Box),g_i)-\Delta(-\Box,g_i)\biggr)d_i \tilde g_i k^{d_i}\right)
\end{equation}

Again we can define a pure cutoff of generalized type I by replacing
$g_i$ by $\gamma_i k^{d_i}$ in $\calr_k$
\begin{equation}
\label{eq:gen_pure_cutoff}
\calr_k(-\Box,\gamma_i k^{d_i})=\Delta(P_k(-\Box),\gamma_i k^{d_i})
-\Delta(-\Box,\gamma_i k^{d_i})\ .
\end{equation}
The beta functional that one obtains with this cutoff has the form
\begin{eqnarray}
\label{rhs6}
\beta&=&\frac{1}{2}\STr
\left(\Delta(P_k(-\Box),\gamma_i k^{d_i})+\Delta(-\Box,g_i)
-\Delta(-\Box,\gamma k^{d_i})\right)^{-1}
\times
\nonumber\\
&&
\left(\frac{\partial\Delta}{\partial(-\Box)}
\frac{\partial \calr_k(-\Box)}{\partial t}
+\left(\frac{\partial\Delta}{\partial g_i}(P_k(-\Box),\gamma_i k^{d_i})
-\frac{\partial\Delta}{\partial g_i}(-\Box,\gamma_i k^{d_i})
\right)
d_i \gamma_i k^{d_i}\right)
\end{eqnarray}

Again we see that the two beta functionals have the same form except for
the replacement of $\tilde g_i$ by $\gamma_i$ in certain functional dependences;
additional terms in the first factor cancel for $\gamma_i=\tilde g_i$.
Therefore the argument given above shows that if we set $\gamma_i=\tilde g_{i*}$
the pure cutoff will have a FP in the same position as the generalized type I cutoff.

The reason why this discussion is less general than the previous one is that this
type of cutoff could only be defined if the inverse propagator has a specific form.
It may be possible to generalize this argument, for example defining the cutoff by the rule
$$
\nabla_\mu\ \mapsto \ \sqrt{\frac{P_k(-\Box)}{-\Box}}\,\nabla_\mu\ .
$$
We will not pursue this further.
The discussion of the type III cutoff is sufficient to make the point in generality.
Furthermore, the type III cutoff is ``ideologically'' at the opposite extreme
of a pure cutoff, being always fully dependent on all couplings.
This is also supported by the numerical results, which show that type III
cutoffs yield fixed points at at the extreme end of the range of variation \cite{CPR2}.
So it is somewhat reassuring that one can reproduce at least the FP position
of a spectrally adjusted, type III cutoff by a pure cutoff.

\bigskip
\goodbreak

\centerline{Acknowledgements}
R.P. would like to thank M. Reuter for discussions on cutoff choices.


\begin{thebibliography}{10}

\bibitem{Reuter}
M. Reuter, Phys. Rev. {\bf D57}, 971 (1998) [arXiv:hep-th/9605030].
%
\bibitem{Dou}
D.~Dou and R.~Percacci,
Class.\ Quant.\ Grav.\  {\bf 15} (1998) 3449; [arXiv:hep-th/9707239].
%
\bibitem{Souma} 
W. Souma, Prog. Theor. Phys. {\bf 102}, 181 (1999); [arXiv:hep-th/9907027].
%
\bibitem{Lauscher} 
O. Lauscher and M. Reuter, Phys. Rev. {\bf D65}, 025013 (2002);
[arXiv:hep-th/0108040];
Class. Quant. Grav. {\bf 19}, 483 (2002);
[arXiv:hep-th/0110021];
Int. J. Mod. Phys. {\bf A 17}, 993 (2002);
[arXiv:hep-th/0112089];
M. Reuter and F. Saueressig, Phys. Rev. {\bf D65}, 065016 (2002).
[arXiv:hep-th/0110054].
%
\bibitem{Litim}
D.~F.~Litim, Phys.Rev.Lett. {\bf 92} 201301 (2004);
P.~Fischer and D.~F.~Litim,
Phys.\ Lett.\  B {\bf 638} (2006) 497 [arXiv:hep-th/0602203].
%
\bibitem{Eichhorn}
  A.~Eichhorn, H.~Gies and M.~M.~Scherer,
  arXiv:0907.1828 [hep-th].




\bibitem{Lauscher2}
O. Lauscher and M. Reuter, Phys. Rev. {\bf D 66}, 025026 (2002), [arXiv:hep-th/0205062].
%
\bibitem{Codello}
A. Codello and R. Percacci, Phys.Rev.Lett. {\bf 97}, 221301 (2006) arXiv: hep-th/0607128.
%
\bibitem{BMS}
  D.~Benedetti, P.~F.~Machado and F.~Saueressig,
Mod. Phys. Lett. {\bf A24} 2233-2241 (2009) arXiv:0901.2984 [hep-th];
Nucl. Phys. {\bf B824} 168-191 (2010) arXiv:0902.4630 [hep-th].
%
\bibitem{Niedermaier:2009zz}
  M.~R.~Niedermaier,
  Phys.\ Rev.\ Lett.\  {\bf 103}, 101303 (2009).
%
\bibitem{CPR1}
A.~Codello, R.~Percacci and C.~Rahmede,
Int.\ J.\ Mod.\ Phys.\  A {\bf 23} (2008) 143 [arXiv:0705.1769 [hep-th]]; 
%
\bibitem{CPR2}
 A.~Codello, R.~Percacci and C.~Rahmede,
  Annals Phys.\  {\bf 324} (2009) 414
  [arXiv:0805.2909 [hep-th]].
%
%
\bibitem{MachSau}
  P.~F.~Machado and F.~Saueressig,
  Phys.\ Rev.\  D {\bf 77}, 124045 (2008)
  [arXiv:0712.0445 [hep-th]].
%
\bibitem{PercacciN} 
R. Percacci, 
Phys. Rev. {\bf D73}, 041501(R) (2006);
[arXiv:hep-th/0511177].
%
\bibitem{conformal}
  M.~Reuter and H.~Weyer,
  arXiv:0801.3287 [hep-th]; 
  Phys.\ Rev.\  D {\bf 80}, 025001 (2009)
  [arXiv:0804.1475 [hep-th]].
M.~Reuter and H.~Weyer, 
 Phys.\ Rev.\  D {\bf 79} (2009) 105005 and
\mbox{arXiv:0801.3287 [hep-th]};
 Gen.\ Rel.\ Grav.\  {\bf 41} (2009) 983 and 
\mbox{ arXiv:0903.2971 [hep-th]}.
  P.~F.~Machado and R.~Percacci,
  Phys.\ Rev.\  D {\bf 80}, 024020 (2009)
  [arXiv:0904.2510 [hep-th]].
%
\bibitem{Weinberg} S. Weinberg, In {\it General Relativity: An Einstein centenary survey},
ed. S.~W. Hawking and W. Israel, pp.790--831;
Cambridge University Press (1979).
%
%
\bibitem{AS_rev} 
Max Niedermaier and Martin Reuter,
Living Rev. Relativity 9,  (2006), 5; 
M.~Niedermaier, Class.\ Quant.\ Grav.\  {\bf 24} (2007) R171 [arXiv:gr-qc/0610018];
R. Percacci, ``Asymptotic Safety'', 
in ``Approaches to Quantum Gravity: Towards a New Understanding of Space, Time and Matter'' 
ed. D. Oriti, Cambridge University Press (2009) arXiv:0709.3851 [hep-th];
D.~F.~Litim,
arXiv:0810.3675 [hep-th].
%
\bibitem{Wetterich}
C. Wetterich, Phys. Lett. {\bf B 301}, 90 (1993).
%
\bibitem{Buchbinder}
I.L. Buchbinder, S.D. Odintsov and I.L. Shapiro,
``Effective action in quantum gravity'',
IOPP Publishing, Bristol (1992).

\bibitem{elisa1}
  E.~Manrique and M.~Reuter,
  Phys.\ Rev.\  D {\bf 79}, 025008 (2009)
  [arXiv:0811.3888 [hep-th]];

\bibitem{elisa2}
E.~Manrique and M.~Reuter,
arXiv:0907.2617 [gr-qc].

\bibitem{Perini3}
  R.~Percacci and D.~Perini,
  Class.\ Quant.\ Grav.\  {\bf 21}, 5035 (2004)
  [arXiv:hep-th/0401071].

\bibitem{gies}
J.~M.~Pawlowski,
Int.\ J.\ Mod.\ Phys.\  A {\bf 16}, 2105 (2001);
Annals Phys.\  {\bf 322}, 2831 (2007) [arXiv:hep-th/0512261];
H. Gies, Phys. Rev. {\bf D 66} 025006 (2002) arXiv:hep-th/0202207

%
\bibitem{optimized}
  D.~F.~Litim,
  Phys.\ Lett.\  B {\bf 486} (2000) 92
  [arXiv:hep-th/0005245].
  Phys.\ Rev.\  D {\bf 64} (2001) 105007
  [arXiv:hep-th/0103195].

\bibitem{wegner}
F.J. Wegner, J. Phys. {\bf C 7} 2098 (1974).
%
\bibitem{Perini1} 
R. Percacci and D. Perini, Phys. Rev. {\bf D67}, 081503(R) (2003) [arXiv:hep-th/0207033];
Phys.~Rev. {\bf D68} (2003) 044018 [arXiv:hep-th/0304222].

\end{thebibliography}
\end{document}